\documentclass[12pt,preprint]{aastex}
\usepackage{graphicx}
\usepackage{ulem}
\usepackage{longtable,lscape}

\def\ie{\mbox{i.e.}}
\def\eg{\mbox{e.g.}}

\def\teff{\mbox{T$_{\rm eff}$}}
\def\logg{\mbox{log\,{\it g}}}
\def\vmicro{\mbox{$\xi_{\rm t}$}}
\def\kmsec{\mbox{km~s$^{\rm -1}$}}
\def\loggf{\mbox{$\log gf$}}

\def\vsini{\mbox{$v\,{\rm sin}{\it i}$}}

\begin{document}
\title{Horizontal-branch morphology and multiple stellar populations in the anomalous globular cluster M\,22.
\thanks{Based on data collected at the European 
Southern Observatory with the FLAMES/UVES spectrograph under the program 085.D-0698A.}}
\author{
 A.\ F. \,Marino\altaffilmark{2,3},
 A.\ P. \,Milone\altaffilmark{4,5,3}, and
 K.\, Lind\altaffilmark{2} 
 }

\altaffiltext{2}{Max Planck Institute for Astrophysics, Postfach 1317,
	   D-85741 Garching, Germany; amarino@MPA-Garching.MPG.DE}

\altaffiltext{3}{Research School of Astronomy \& Astrophysics, Australian National University, Mt Stromlo Observatory, via Cotter Rd, Weston, ACT 2611, Australia}

\altaffiltext{4}{Instituto de Astrof\`\i sica de Canarias, E-38200 La
              Laguna, Tenerife, Canary Islands, Spain; milone@iac.es}

\altaffiltext{5}{Department of Astrophysics, University of La Laguna,
           E-38200 La Laguna, Tenerife, Canary Islands, Spain}

\shorttitle{HB morphology and stellar populations in M\, 22} 
\shortauthors{Marino et al.\ } 

\begin{abstract}
M\,22 is an anomalous globular cluster that hosts two groups of stars with different
 metallicity and $s$-element abundance. The star-to-star light-element variations in both groups, with the presence of individual Na-O and C-N anticorrelations, demonstrates 
that this Milky-Way satellite has experienced a complex star-formation history.
We have analysed FLAMES/UVES spectra for seven stars covering a small color interval, on the reddest horizontal-branch (HB) portion of this cluster and investigated
possible relations between the chemical composition of a star and its location along the HB. 
Our chemical abundance analysis takes into account effects introduced by deviations from the local-thermodynamic equilibrium (NLTE effects), that are significant for the measured spectral lines in the atmospheric parameters range spanned by our stars.
We find that all the analysed stars are barium-poor and sodium-poor, thus supporting the idea that the position of a star along the HB is strictly related to the chemical composition, and that the HB-morphology is influenced by the presence of different stellar populations.
\end{abstract}


\section{Introduction}
\label{introduction}
The horizontal-branch (HB) morphology in globular clusters (GCs) is mainly governed by metallicity (e.g.\ Arp et al.\ 1952), which therefore is considered as the {\it first parameter} regulating the HB stars position along the color-magnitude diagram (CMD). 
The more metal-poor the GC, the more extended is the HB.
The so called ``second-parameter problem'' recognises that 
the reality is more complex
as some GCs with similar metallicity exhibit different HBs. Hence, at least one second parameter is required to properly caracterize the HB morphology in GCs (e.g.\ Sandage \& Wallerstein 1960, van der Berg 1965).

While the second parameter has been widely studied since the sixties, its identity is still unclear. Stellar mass, rotation, cluster central density, cluster mass, helium abundance, and age, among others, have been all suggested as possible solutions but none of them is fully satisfactory and possibly more than one parameter is at work (see Catelan 2009, Dotter et al.\ 2010 and references therein).  

It is now widely accepted that nearly all the {\it normal} GCs host stellar populations with different 
light-elements abundance (Carretta et al. 2009, and references therein).   
Observational studies have shown that in GCs co-exist stars belonging to a first generation, whose O and Na composition is compatible with field stars at similar metallicities, and to second generation(s),  \eg\ those manifesting a chemical pattern (O-depletion and Na-enhancement) compatible with a formation from material that underwent  H-burning at high temperatures.
Since the same material depleted in O and enhanced in Na is expected to be also He-enhanced, these findings
provide new insight to the second-parameter problem, and seem to support 
the pioneering studies by Norris \& Freeman (1982)  suggesting that He abundances may take a role in the distribution of stars along the HB of GCs
(e.g.\ D'Antona et al.\ 2005, Lee et al.\ 2005, Piotto et al.\ 2005, 2007, D'Antona \& Caloi 2008).

In this context, the GC M\,4 could be regarded as the {\it Rosetta stone} to connect multiple populations observed in the CMD with the HB morphology. 
This cluster indeed exhibits a strong dichotomy in the abundances of sodium and oxygen 
with the two main populations of Na-poor/O-rich first-generation, and Na-rich/O-poor second-generation stars defining two separate sequences along the red-giant branch (RGB) in the $U$ versus $U-B$ CMD. 
Due to the large differences in the CN and CH bands affecting the ultraviolet region of the spectrum, Na-rich stars define a sequence on the red side of the RGB, while Na-poor stars populate a bluer, more spread sequence (Marino et al.\ 2008).
Interestingly, M\,4 hosts a bimodal HB, which is well populated
both on the red and the blue side of the RR-Lyrae gap. 
It turns out that HB stars exhibit a bimodal Na and O distribution similar to that found along the RGB, with red-HB stars having solar [Na/Fe] and blue-HB ones being O-depleted and Na-enhanced (Marino et al.\ 2011a, Villanova et al.\ 2012). 
The position of first-generation HB stars, whose O and Na composition is compatible with field stars at similar metallicities, populate the red-HB. On the other hand, second-generation stars, supposed to form 
from material that underwent  H-burning at high temperatures, 
distribute on the blue HB (see also Norris et al.\ 1981 and Smith \& Norris 1993).

A similar scenario applies for other GCs.
Villanova et al.\ (2009) analysed spectra of seven HB stars with effective temperatures 8000$<\teff<$9000~K in NGC~6752 and found that six of them have a chemical content that resembles the field-halo composition, \ie\ being the progeny of first stellar generation identified on the RGB.
Successively, the connection between stellar populations and HB morphology in GCs  has been confirmed for
NGC\,2808 (Gratton et al.\ 2011), NGC\,1851 (Gratton et al.\ 2012a), 47\,Tucanae (Milone et al.\ 2012a, Gratton et al.\ 2012b), NGC\,6397 (Lovisi et al.\ 2012), and M\,5 (Gratton et al.\ 2012b).

While the majority of GCs are almost homogeneous in iron,
recent studies reveal that there exists a group of anomalous GCs (AGCs), characterised by internal variations in the bulk heavy-element content, including iron and elements associated to slow neutron-capture ($s$-elements).
 The chemical analogies with the two {\it anomalous} cases $\omega$\,Centauri and M\,54, which are usually associated to the remnants of a dwarf galaxy cannibalised by the Milky Way, suggest the fascinating idea that  AGCs could be the survived nuclei of more massive systems.
 
M\,22 is certainly the prototype of an AGC.
Indeed it hosts two groups of stars with different content of iron and $s$-elements, and this bimodality is associated to a different C+N+O content (Marino et al.\ 2009, 2011b, 2012, Da Costa et al.\ 2009, Alves Brito et al.\ 2012). 
Moreover, each individual group of $s$-rich and $s$-poor stars exhibits large star-to-star light-elements variation, with the presence of Na-O and C-N anticorrelations (Marino et al.\ 2011b). 

In this paper we analyse the chemical abundances for seven HB stars of M\,22 and investigate the possibility that stellar populations previously identified in this cluster could be related to a different position along the HB.

\section{Observations and data analysis}
\label{data}

Our dataset consists of UVES spectra collected within a project devoted to the study of sub-giant stars in M22 with GIRAFFE, 
(see  Marino et al.\ 2012).
The fibers feeding the UVES spectrograph were centered on seven HB stars distributed along the more luminous portion of the blue HB of M22.
The UVES spectrograph was used in the RED 580 setting providing spectra of R$\sim$45000. The spectra have a spectral coverage of $\sim$2000~\AA\ with the central wavelength at 5800~\AA. 
All our target stars were observed in the same FLAMES configuration in four different exposures of 46
minutes plus one exposure of 26 minutes, for a total observing time
of 210 minutes. The single exposures were merged together, after they were reportad  at the rest wavelength.
The typical signal to noise ratio of the combined spectra is in the range 70$<S/N<$100.
Data were reduced using UVES pipelines (Ballester et al.\ 2000), including bias subtraction, flat-field correction, wavelength calibration and sky subtraction.

We derived radial velocities (RVs) by cross-correlating the observed individual exposures with a spectrum simulated with atmospheric parameters close to our HB targets. The observed/template spectrum matches, after the heliocentric
correction, yielded for our sample a mean RV of $-144\pm2$~\kmsec\
(rms=6~\kmsec). This value is in reasonable agreement with the values in the literature (e.g.\,
$-143\pm1$~\kmsec, rms=9~\kmsec from Marino et al.\ 2012; $-146.3\pm0.2$~\kmsec, rms=7.8~\kmsec, from the value reported in the Harris catalog).
The rms/$\sqrt{(N-1)}$ (where $N$ is the number of available exposures) of our RVs measurements for each star is $\sim$0.6 \kmsec.
This value has been taken as an estimate of the internal error associated with our RV values. 
The RVs derived from different exposures of the same star agree within our errors.  
Hence, as our spectra were collected in a time interval of $\sim$3 months, the presence of binaries with period up to $\sim$150-200 days can be excluded.

The RVs of all the observed stars lie within $\pm$3~$\sigma$ of the average RV (see Tab.~\ref{data}), that is the condition required for a star to be cluster member. 
However, in the case of M22, the RV-based criteria used to assess the cluster membership may be  weak. Indeed M22 is projected toward the bulge, and the field is kinematically complex.
Since the probability of finding a 
{\it very} metal-poor field star with kinematics compatible with the cluster is extremely low, the metallicities obtained for our stars (listed in Tab.~\ref{data}, see Sect.~\ref{atm}), all consistent with the M22 metallicities  (\eg\ Marino et al. 2009), constitute a more efficient tool to clean our sample from field contaminants.
The status of He-burning star, as indicated by the derived atmospheric parameters (see Sect.~\ref{atm}) not compatible with a
main sequence star (except maybe for one target), can be used as an additional proof, because
the probability of a field contaminant with such properties is even lower.
So, we conclude that all our observed stars are cluster members.

We provide an estimate of the projected rotational velocities (\vsini) for our stars by  deconvolving the full width at half
maximum of the cross-correlation profile for the widening of the synthetic template computed with the instrumental profile of our data.
Although 
a study of \vsini\ is beyond the aims of this work, our values provide an estimate of the additional broadening to be applied to the lines when using spectral synthesis (see Sect.~\ref{atm}).
The \vsini\ of our sample stars ranges from 12 to 31~\kmsec\ in full agreement with values determined
for HB stars with similar \teff\, in a sample of other GCs  (e.g.\ Recio-Blanco et al. 2004; Behr et al. 2000).

\begin{figure}[ht!]
 \centering
 \epsscale{.75}
    \plotone{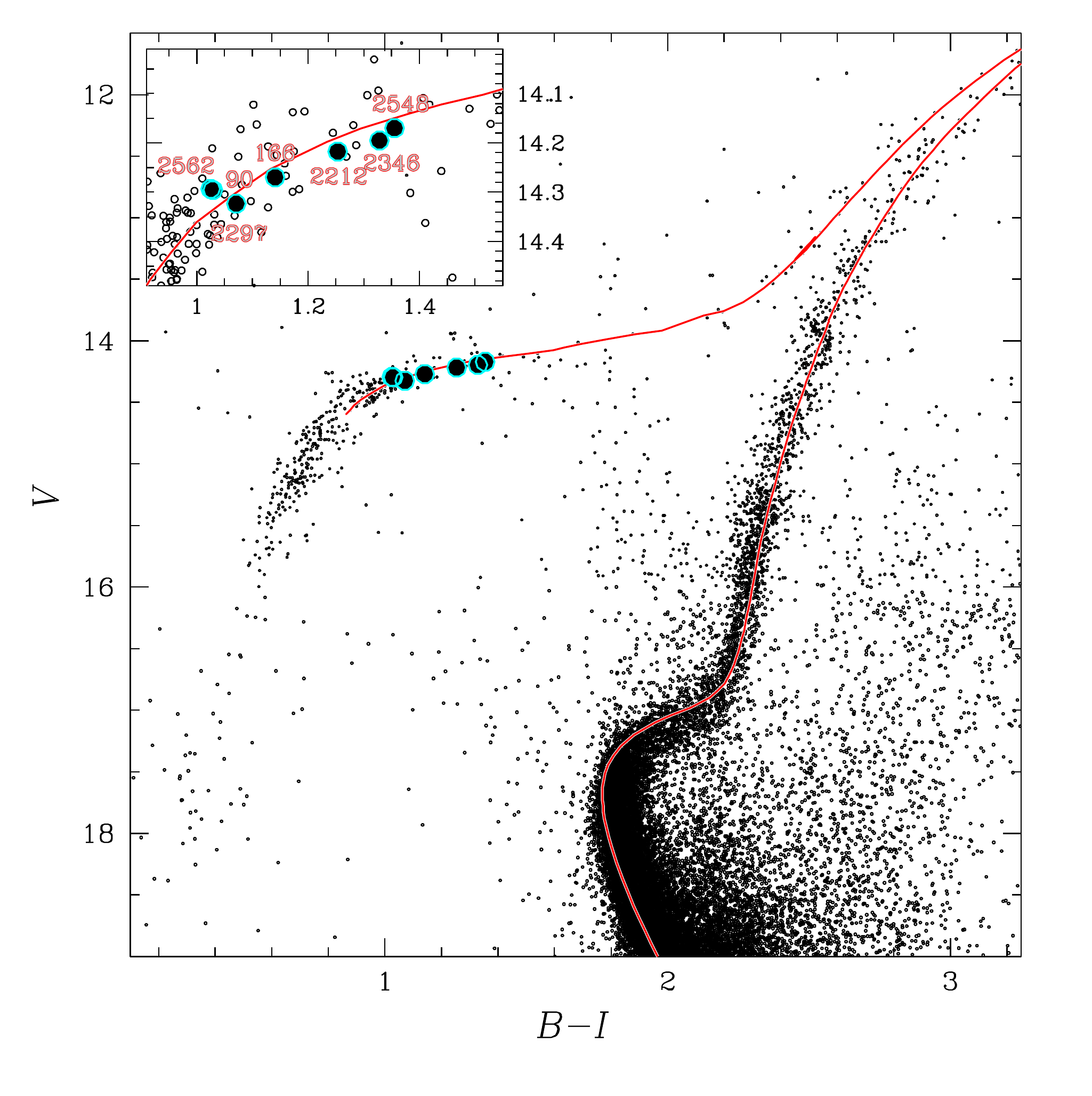}
   \caption{ $V$ versus $B-I$ CMD from P. Stetson of M\,22 corrected for differential reddening. Spectroscopic targets are marked with circles. We superimposed to the CMD the best-fitting isochrone from the Padova database (red continuous line), and an isochrone corresponding to 1.8$M_{\rm \odot}$ star (blue dotted line). The inset is a zoom around the HB. }
       \label{cmd}
 \end{figure}

Our photometry included Johnson $B$, $V$, $I$ bands from Peter Stetson's database\footnote{\sf http://www3.cadc-ccda.hia-iha.nrc-cnrc.gc.ca/community/STETSON/index.html}. Photometric catalogs were corrected for differential reddening as in Milone et al.\ (2012b) and were already used in  Marino et al.\ (2011b), to which we refer the reader for details.
The $V$ versus $B-I$ color-magnitude diagram (CMD) corrected for differential reddening is shown in Fig.~\ref{cmd}, where we have marked the spectroscopic targets with cyan dots. 

Alpha-enhanced isochrones from the BaSTI and the Padova database (Pietrinferni et al.\ 2004, Girardi et al.\ 2002) have been used to fit the observed CMD.
We assumed reddening E$(B-V)$=0.35, [Fe/H]=$-$1.85, and distance 
modulus $(m-M)_{\rm V}$=13.65, in agreement with values in the literature (Harris 1996, updated as in 2010). 

\subsection{Atmospheric parameters\label{atm}}
Spectroscopic measurements were conducted with the local thermodynamic equilibrium (LTE) code MOOG (Sneden 1973) and by using model atmospheres interpolated from the grid of Castelli et al.\ (2004).
In general, except for the cooler stars, too few Fe lines could be measured in our spectra to properly constrain all the atmospheric parameters.
Therefore, we derived gravities (\logg) from the isochrone that best fit our CMD, and associated to each star the \logg\ value corresponding to the point on the isochrone with the smallest $distance$ from the star. The $distance$ is computed as in Gallart et al.\ (2003, Sect.~4) by enhancing the difference in color by a factor of seven with respect to the difference in magnitude. 
As pointed out by Gallard and collaborators,
the factor seven, that is determined empirically, provides a weight to the color of a star that is larger than that of the magnitude. This takes into account the fact that, for a given isochrone, a star has a color that is better constrained than the magnitude, as any uncertainty in distance, gravity, or reddening, corresponds to a magnitude variation that is larger than that in color.

Once \logg\ had been fixed, effective temperatures (\teff) and  microturbolence values (\vmicro) were derived from the spectra, by satisfying the ionisation equilibrium between Fe~I and Fe~II abundances, and by removing trends in Fe abundances{\footnote {Fe~I or Fe~II, depending on the number of lines available for the different ionisation stages.}} with reduced equivalent width (EW), respectively. 
We have used the non-LTE (NLTE) spectral code MULTI (Carlsson 1986; see Lind et al.\ 2012 and Bergemann et al. 2012, for a description of the NLTE modelling) to estimate the offset between Fe~I and Fe~II abundances expected for each star.
It turns out that in LTE approximation the abundances from Fe~I lines are lower than the ones from Fe~II lines by $\sim$0.25-0.28 dex, due mainly to large positive NLTE effects on FeI lines.
This offset has been considered in the ionisation balance for the spectroscopic determination of \teff. 
Our approach has the important advantage that it is independent from photometry and is not affected by the differential reddening effects present across the face of M22.
On the other hand, \logg\ is more robustly internally fixed from isochrones, given that our stars distribute on an almost horizontal section of the HB and are expected to show similar gravities.
For one star (\#2562) no Fe~I line could be measured from the spectrum, and we adopted the \teff\ derived from the best-fitting isochrone. 

The final adopted atmospheric parameters for each star are listed in Tab.~\ref{data}. 
Internal errors in \logg\ and \teff\ depend on photometric uncertainties and the ionisation equilibrium, respectively.
For our relatively bright stars, internal errors in the color $(B-I)$ and in the magnitude $V$, including uncertainties due to the differential reddening correction, are 0.025 and 0.015 mag respectively (Milone et al. 2012b). 
A shift of our target stars in the position along the ZAHB by these photometric uncertainties corresponds to a variation in \logg\ of  $\sim$0.02, that is so small to translate in negligible internal errors in the derived chemical abundances (see Sect.~\ref{abund}).
To determine internal errors in temperatures, we adopted the following approach to all the stars in our sample: {\it (i)} we added the standard error of the mean for Fe~I and Fe~II in quadrature; {\it (ii)} then adjusted \teff\ until the quantity Fe~I$-$Fe~II, after the NLTE corrections,
was equal to this value. 
The mean difference between the new \teff\ and the original values is $\sim$170 K, that is an estimate of the internal uncertainties affecting our \teff\ determinations. 
For \vmicro\ we measured
the formal uncertainty in the slope between the Fe abundances and the reduced EWs, and adjusted until the formal slope was equal to this value. Given the relatively low number of spectral lines available, internal uncertainties in \vmicro\ are large, and are on average equal to $\sim$0.5 \kmsec.

For a comparison between HB and RGB chemical abundances, that is the subject of our study, an estimate of possible systematic errors affecting the model atmospheres is required. 
External errors are much harder to estimate. A comparison of atmospheric parameters derived with different techniques may be indicative of possible systematics affecting our adopted values.
In Tab.~\ref{data}, along with the adopted parameters we also show the \teff\ determined from isochrones, and \vmicro\ resulting from the \teff-\vmicro\ empirical relation 
derived in Pace et al. (2006)  from the least square fit of the HB data analysed by Behr et al. (2000).
A comparison of the two sets of \teff\ suggests that the agreement is satisfactory for the three coldest stars (\eg\ \#2212, \#2346, and \#2548), whose difference among spectroscopic and photometric \teff\ is at most $\sim$100~K.
For the two hottest stars (\#2297 and \#90) spectroscopic \teff\ are significantly lower, suggesting that 
a possible systematic between the two sets of \teff\ values 
could be higher for warmer stars.
On average the difference between the two sets of \teff\ values (\#166 and \#2562 have been excluded in this calculation) is \teff$_{\rm , adopted}-$\teff$_{\rm , isochrones}$=$-174\pm90$ K,  that could be taken as a rough estimate of possible systematics affecting our \teff\ determinations. We note that,
although the limited number of stars presented here do not allow us to provide a more exhaustive
analysis of possible systematics, we consider the spectroscopic \teff\ more reliable since the photometric ones are inconsistent with the Fe ionisation balance in NLTE. 
The difference between the adopted \vmicro\ and the ones obtained by the Pace et al. relation is \vmicro$_{\rm , adopted}-$\vmicro$_{\rm , Pace}$=$-0.28\pm0.13$ \kmsec. 
To estimate the error provided by the adopted stellar models, we compared the values of \teff\ and \logg\ from the two independent set of isochrones used in this paper. Differences in temperatures and gravities are typically  $\sim$80\,K and $\sim$0.04 dex, respectively, with the BaSTI isochrones providing smaller values of \teff\ and \logg. 
In the following, we will use the \logg\ provided by the Padova isochrones only. 

Our gravities values are based on the assumption that our stars lie on the ZAHB.
During the He-burning phase, however, the stars are expected
to decrease their gravity, which are about 0.2 dex lower for terminal-age HB stars.
The photometric changes during this evolution should be negligible. Hence, \logg\ values derived from isochrones could be overestimated by up to $\sim$0.2 dex if the target is close to the end of its HB life. 
Moreover, canonical models with no He-enhancement, as those we used here, predict \logg\ higher than \logg\ from He-enriched models. 
In our models gravities derived from isochrones at Y=0.30 are higher by $\sim$0.25 dex than those derived by assuming primordial He.
Some evidence for gravities systematically lower by 0.3 dex than those predicted by stellar models, have been shown in the case of $\omega$~Cen (\eg\ Moni Bidin et al. 2011, Moehler et al. 2011), but not in {\it normal} GCs (Moni Bidin et al. 2007, 2009).
As M\,22 shares many chemical features with the most extreme GC $\omega$~Cen (\eg\ Marino et al. 2009; Da Costa \& Marino 2011), 
the current models could be similarly inadequate to describe HB stars in M\,22.
For these reasons, gravity estimates independent on isochrones are important to have an idea of possible systematics in \logg\ introduced by these effects.

\begin{figure}[ht!]
 \centering
 \epsscale{0.5}
    \plotone{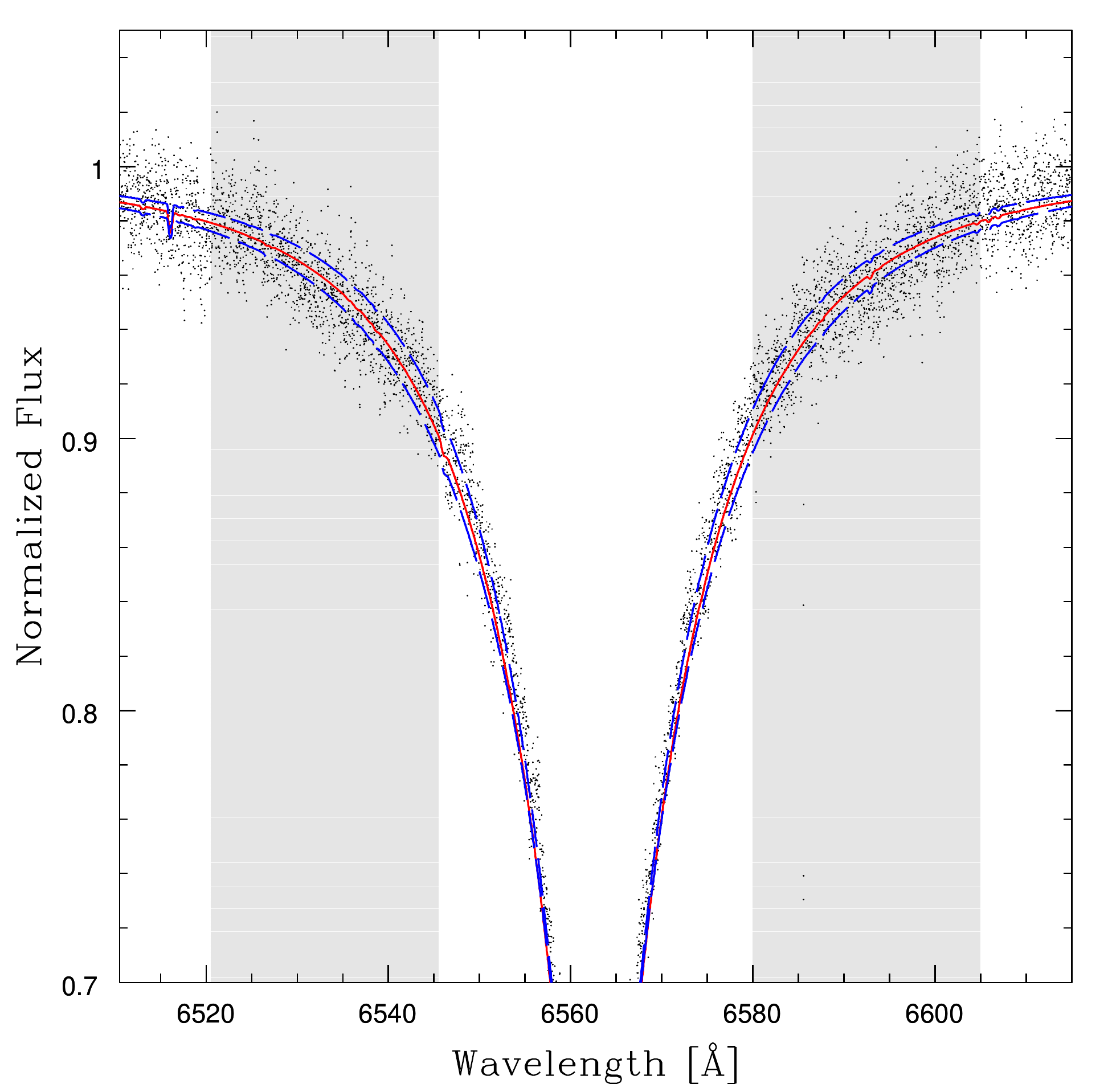}
   \caption{Observed spectrum around the $\rm {H_{\alpha}}$ region of the star \#90. The red spectrum is the model that best-fits the data (\logg=3.13), while the blue dashed spectra are the models with \logg\ higher and lower by $\pm$0.30 dex than the best-fit value. The grey regions are where the minimum $\chi^{2}$ has been determined.}
       \label{Halpha}
 \end{figure}

Two Balmer lines lie in the spectral range of our data: the H$_{\rm\alpha}$ ($\sim$6563~\AA) and the H$_{\rm\beta}$ ($\sim$4861~\AA). 
As, in the \teff\ range of our stars, the hydrogen lines are not very sensitive to the temperature, we can provide  estimates of  gravity from a $\chi^2$ minimisation of the observed line wings with theoretical spectra. To this aim we calculated spectral models for the adopted \teff\ (Tab.~\ref{data}) using the spectral synthesis code SYNTHE\footnote{{\sf http://wwwuser.oat.ts.astro.it/castelli/sources/synthe.html}}. 
Internal errors associated with these measurements are strongly dominated by normalisation uncertainties, that are much larger in the case of  the H$_{\rm\beta}$ lying in the proximity of the blue-edges of the spectra.
For this reason, we preferred to use only the H$_{\rm\alpha}$ for the \logg\ estimates. 
As an example, in Fig.~\ref{Halpha} we represent the spectral synthesis for the star \#90.
Together with the model providing the minimum $\chi^2$, at \logg=3.13, we plotted the models
at \logg=$\pm$0.30 dex around the best-fitting model. The fit with synthetic spectra at different \logg\ suggests that,  due to the quality of our observed spectra and the sensitivity of the H$_{\rm\alpha}$ line wings to \logg, internal errors associated with these measurements are $\sim$0.15-0.20 dex.
The \logg\ values derived from the H$_{\rm\alpha}$ best-fit models, listed in Tab.~\ref{data}, are all (apart for the star \#2562) lower than those derived from isochrones. On average the difference among the gravities obtained from the two different techniques is $-0.14\pm0.08$ dex (rms=0.17), that is an estimate of possible systematics affecting our \logg\ determinations.
We verified that lowering gravities by this quantity 
reflects in temperatures lower by $110\pm8$ K needed to establish the ionisation equilibrium.

In the following, we will use \logg\ provided by isochrones, because they provide a higher level of internal precision, and \teff\ from the ionisation equilibrium. 
For comparison purposes, atmospheric parameters obtained with different techniques are listed in  Tab.~\ref{data}. As discussed, our analysis suggests that the offsets with respect to the adopted parameters are at most of $\sim$200 K in \teff,  $\sim$0.15 dex in \logg, and $\sim$0.30 \kmsec\ in \vmicro. The impact of these possible systematics on chemical abundances is discussed in Sect.~\ref{abund}.

Note that for one star in our sample (\#166) the model atmosphere entirely derived from isochrones gives 
a large difference (0.68 dex) among Fe~I and Fe~II abundances. 
By deriving \teff\ and \vmicro\ from the Fe lines, once \logg\ has been fixed from isochrones (as done for the other stars), we need to significantly lower \teff\ and \vmicro\ to 7450 K and 1.0 \kmsec, respectively. While \teff\ obtained from the ionisation equilibrium is comparable with the lower-\teff\ stars in our sample, \vmicro\ is too low. Once fixed the other parameters, gravity from H$_{\rm\alpha}$ results slightly higher than the isochrone value (see Tab.~\ref{data}).
The spectrum of this star shows a relatively sufficient number of Fe~I (19) and Fe~II (4) lines to attempt to entirely constrain parameters from spectral lines, \eg\ to derive both \teff\ and \logg\ we impose the excitation and ionisation equilibrium, respectively. 
The \logg\ value that establishes the ionisation balance of iron lines, after the offset between Fe~I and Fe~II lines expected in the LTE approximation has been considered, is almost one dex higher than that provided by isochrone fitting, while we still get \vmicro\ values lower than in the other stars.
The reason for this discrepancy is unclear. 
Magnitude and color provided for star \#166 by a different photometric catalog (from Monaco et al.\ 2004) are very close to those of this paper, thus suggesting that the high value of \logg\ is physical. 
One possibility could be that \#166 is an evolved Blue Straggler. Indeed, the comparison of the CMD of Fig.~\ref{cmd} with an 1.1 Gyr isochrone with the same metallicity as M\,22 shows that the color and magnitude of \#166  matches the location of an evolved 1.8 $M_{\rm \odot}$ star. 
We leave this issue to future investigations. Here, we use fully spectroscopically derived atmospheric parameters for this star, and suggest caution in considering its derived chemical abundances.

\subsection{Chemical abundances\label{abund}}
Besides a few Fe~I and Fe~II absorption lines, our spectra show the Na resonance doublet at $\sim$5890~\AA, and few lines of Mg, Ca, Ti, Ba, and Cr. Line broadening from stellar rotation is evident in the spectra, but even in the most extreme cases, the line profiles were close to Gaussian, so line EWs were measured by least-square fitting of Gaussian profiles to the data. 
Adopted \loggf\ are from the NIST database\footnote{{\sf http://physics.nist.gov/PhysRefData/ASD/lines\_form.html}} for all the lines, except for Fe~II for which we used the  \loggf\ from Mel{\'e}ndez \& Barbuy (2009).  

In the five (out seven) coolest stars in our sample we detect the two Ba lines at $\lambda \sim$4934~\AA, and $\sim$6141~\AA.  
These two lines were measured by using spectral synthesis, to properly account for the blends with Fe lines affecting both lines and the effects of hyperfine-splitting.
For the synthesis we used linelists based on Kurucz line compendium\footnote{{\sf http://wwwuser.oat.ts.astro.it/castelli/linelists.html}}, apart from the Ba transition for which we added hyperfine structure and isotopic data from Gallagher et al. (2010). 

As Na resonance spectral lines are heavily affected by departures from LTE, we determined NLTE corrections to our abundances as in Lind et al. (2011), tailored
with our Kurucz model atmospheres.
The overall effect of the NLTE corrections to our abundances is a decrease of the derived [Na/Fe] abundances 
and of the line-by-line scatters.

Our derived chemical abundances are listed in Tab.~\ref{abundances}. 
For Fe and Na we report both the LTE and NLTE abundances.
We note that the NLTE mean metallicity derived for HB stars, that is [Fe/H]$=-1.65\pm0.04$ (rms=$0.10$) agrees within $\sim2\sigma$ with the [Fe/H]  derived on the RGB of the cluster [Fe/H]$_{\rm LTE}=-1.76\pm0.02$ (rms=$0.10$) (Marino et al. 2009).
This agreement is satisfactory if we consider the different spectral lines measured in RGB and HB stars and the different analysis used in Marino et al. (2009, 2011b). 
Indeed, in the atmospheric parameters domain of the M22 RGB stars the NLTE corrections on Fe abundances are much smaller than for HB stars. As such, the ionisation equilibrium between Fe~I and Fe~II is not seriously affected by NLTE effects, and the atmospheric parameters were derived in LTE.
Accounting for the small NLTE effects affecting Fe~I lines in RGB stars, the mean metallicity of M22 would have been systematically higher by $\sim$0.06 dex (Lind et al. 2012), thus decreasing further the HB-RGB mean metallicity offset. 
A more proper comparison between RGB and HB stars in M22 has to take into account the different stellar groups hosted in the cluster, which is the subject of Sect.~\ref{results}.

To have an estimate of internal errors introduced by 
model atmospheres and EW uncertainties in the abundances, we repeated the abundance measurements 
by changing one at a time \teff, \logg\ and \vmicro\ by $\Delta$(\teff)$=\pm$170 K, $\Delta$(\logg)$=\pm$0.02 dex,  $\Delta$(\vmicro)$=\pm$0.50 \kmsec, that are the 
internal uncertainties associated with the atmospheric parameters, as determined in Sect.~\ref{atm}. Metallicity has been changed by 
 $\Delta$([A/H])$=\pm$0.10, as suggested by the derived dispersion in Fe~II abundances, and EWs by $\Delta$(EW)$=\pm$8~m\AA, that is the typical error associated with our measurements as we have verified by comparing EWs obtained from single exposures of the same star.
Internal uncertainties in chemical abundances have been determined by applying this procedure for all the stars in our sample. Then, the average uncertainties for each chemical species, in Tab.~\ref{errors}, have been taken as estimates of the errors in abundances. 
Assuming that the uncertainties listed in Tab.~\ref{errors} are uncorrelated, we estimate total abundance uncertainties by summing in quadrature the various contributions. 
The resulting internal errors in chemical abundances ($\sigma_{\rm total}$ in Tab.~\ref{errors}) are typically of $\sim$0.10-0.20 dex and, in most cases, are comparable to the observed dispersions. 
In Tab.~\ref{errors} we also list the sensitivity of derived abundances to possible systematics that may affect our atmospheric parameters (see Sect.~\ref{atm}).

\section{Results}
\label{results}

\begin{figure*}[ht!]
  \centering
  \epsscale{.85}
     \plotone{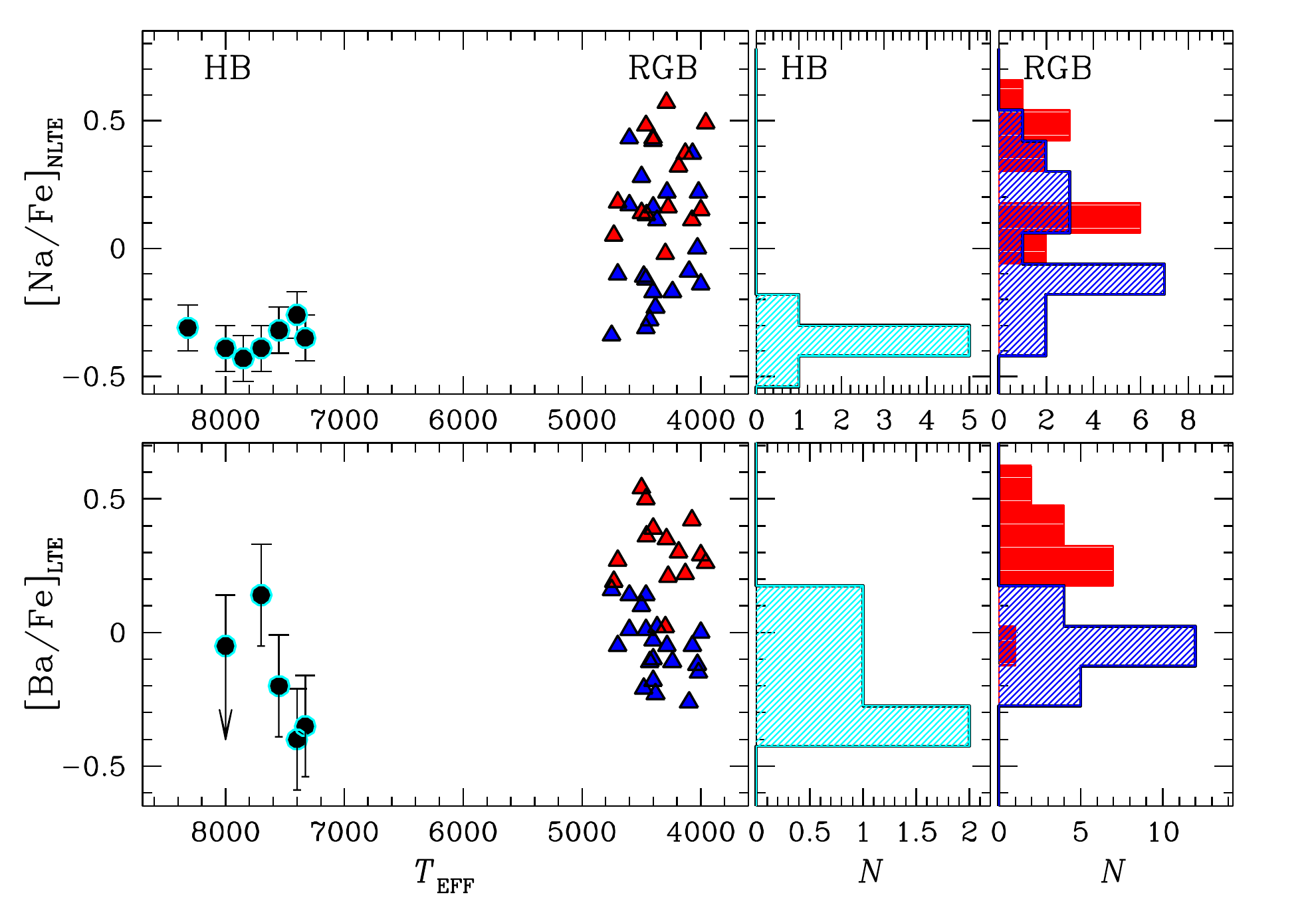}
    \caption{\textit{Left Panels.} [Na/Fe] (top) and [Ba/Fe] (bottom) as a function of \teff\ for HB stars (circles) and RGB stars (triangles). $s$-rich and $s$-poor RGB stars are colored red and blue, respectively. The histogram distribution of [Na/Fe] and [Ba/Fe] for HB stars are shown in the middle panels, while the red and blue shaded-histograms in the right panels correspond to $s$-rich and $s$-poor RGB stars, respectively.  Measurements for RGB stars come from Marino et al.\ (2011b). Here we have applied a correction of $+$0.06 dex to the RGB Fe~I abundances according to the NLTE corrections expected for these stars.
  }
        \label{NaBa}
  \end{figure*}

The lower-left panel of Fig.~\ref{NaBa} shows [Ba/Fe] as a function of \teff\ for the HB stars measured in this paper (circles) and RGB stars studied by Marino et al.\ (2009, 2011b, triangles). 
In these plots a correction of 0.06 dex in Fe has been applied to the RGB stars to take into account NLTE effects, as discussed in Sect.~\ref{abund}.

The lower panels of Fig.~\ref{NaBa} show that for RGB stars, the barium distribution is clearly bimodal. We colored red and blue $s$-rich and $s$-poor stars, respectively, as they have been defined by Marino et al.\ (2009, 2011b). 
These color codes will be used consistently hereafter.
The comparison of the histogram distribution of [Ba/Fe] for HB stars (lower-middle panel) and RGB stars (lower-right panel) reveals that all the analysed HB stars are consistent with being the progeny of $s$-poor stars.
Despite that the errors associated with our HB [Ba/Fe] measurements are larger than the ones associated with RGB values, we note that:
{\it (i)} the star showing the higher Ba abundance is the peculiar star \#166, that we do not consider in the discussion;
{\it (ii)} the mean HB Ba content of our analysed stars is low.
Excluding the two stars \#166 and \#90 (for which we can provide only an upper limit to the Ba abundance), the mean [Ba/Fe] for HB stars is lower than the ones derived for $s$-poor stars on the RGB, but still compatible with this value within a 3$\sigma$ level.
Many factors can explain the discrepancy between the mean HB and $s$-poor RGB Ba abundance, such as the different lines analysed, NLTE effects, and/or systematics in our parameters scale. 
We note that the Ba line $\lambda$4934~\AA, that was not used in the RGB analysis,  gives Ba abundances systematically lower by $\sim$0.15 dex than the line $\lambda$6141~\AA, that was instead measured on RGB stars.
The sensitivity of [Ba/Fe] abundances to systematics in the model atmospheres listed in the three right-most columns of Tab.~\ref{errors}, suggests that, even accounting for the largest systematics, it is unlikely to reconcile the HB mean [Ba/Fe] with the $s$-rich RGB mean abundance.
Therefore we can safely associate all the analysed HB stars to the $s$-poor stellar group.

Additional arguments for the association between the $s$-poor group discovered on the RGB and the analysed HB stars come from the inspection of other chemical abundances.
As discussed in Marino et al. (2011b), 
chemical properties characterising the two stellar groups with different $s$-process abundances include that {\it (i)} they have a different metallicity, with $s$-poor stars showing an overall metallicity $\sim$0.15 dex lower than the $s$-rich stars, {\it (ii)} and 
neither the $s$-rich nor the $s$-poor group is
consistent with a simple stellar population, 
as both groups exhibit significant variations in the abundance of light elements carbon, nitrogen, sodium, oxygen, and aluminium 
delineating individual C-N/Na-O/Na-Al (anti)correlations (see Marino et al.\ 2009, 2012).
Unfortunately, the error on Fe for HB stars is much larger than for RGB stars.
The mean [Fe/H] derived from the HB sample perfectly corresponds to the $s$-rich average metallicity, but it still agrees within a 3~$\sigma$ level with the mean value obtained for the $s$-poor giants.
Therefore, we are not able here to use the relatively small difference in metallicity as a tool to associate the analysed HB stars to one or the other RGB stellar group. This is also because the possible interplay of systematics, such as the different analysed lines, different treatment of NLTE effects,  can easily shift our mean metallicity.

As sodium varies much more than [Fe/H] in the different M\,22 stellar populations, for our HB sample, [Na/Fe] abundances could provide some constraints on the belonging of these stars to the different stellar populations identified on the RGB.
In the upper-left panel of Fig.~\ref{NaBa}, we plot [Na/Fe] against \teff\ for HB and RGB stars. Both $s$-rich and $s$-poor RGB stars span a large range in [Na/Fe] with $s$-rich stars having, on average, higher sodium content.
The histogram distribution of [Na/Fe] for $s$-poor stars is multimodal, with about one half of the stars clustered around  [Na/Fe]$\sim -$0.20 dex (upper-right panel of Fig.~\ref{NaBa}). 
None of the RGB $s$-rich stars have [Na/Fe]$\lesssim$0.
The HB stars analysed in this paper span a much more narrow range in sodium. 
Their mean sodium content is [Na/Fe]=$-$0.34$\pm$0.03 (rms=0.06) and they appear to belong to the same group of $s$-poor/Na-poor stars identified along the RGB. The mean Na abundance relative to Fe for these RGB stars is [Na/Fe]=$-0.13\pm0.03$ (rms=0.09), which would decrease by $\sim$0.06 if we account for NLTE corrections on Fe~I RGB abundances.
Despite the mean Na abundances of the analysed HB stars agrees within 3$\sigma$ with the average value measured on $s$-poor/Na-poor RGB stars, we cannot exclude systematic effects due to the different lines analysed.
In any case, the low mean [Na/Fe] measured for the HB analysed stars provides a further and  more robust  indication that the these stars belong to the $s$-poor group. 
In addition to this, within the $s$-poor group, they constitute the progeny of the Na-poor primordial population.

Within observational errors the chemical abundances of the other analysed elements Mg, Ca, Ti, and Cr do not show evidence for internal variations. The $\alpha$ element abundances of Mg, Ca, and Ti relative to iron are typical of Population II stars, as also observed on the RGB (Marino et al. 2011), while chromium abundances relative to iron are roughly solar-scaled.

\section{Conclusions}

Observational evidence of a connection between the distribution of stars along the HB and their light element chemical content have been shown in the {\it normal} GCs 
M\, 4, 47~Tuc, NGC~2808 and NGC~6397 (Marino et al. 2011, Villanova et al. 2012, Gratton et al. 2011, Gratton et al. 2012, Lovisi et al. 2012).
We have analysed spectra for seven stars in M\,22 spanning a small range of temperature (7300$\lesssim\teff \lesssim$8300) along the HB, with the main purpose of measuring their Ba and Na abundances.
Our results support the idea that the distribution of stars along the HB is somehow related to the multiple stellar population phenomenon also for the {\it anomalous} GC M\,22. 

Recent studies on RGB and SGB stars have shown that M\,22,  differently from the most ({\it normal}) GCs,
hosts two groups of stars with different content of iron, $s$-elements, and overall C+N+O (Marino et al. 2009, 2011, 2012; Da Costa et al.\ 2009, Alves Brito et al.\ 2012).
Both the iron-/$s$-/CNO-rich and iron-/$s$-/CNO-poor group exhibits 
an internal scatter in the abundance of light elements (e.g.\, C, N, O, Na, Al) with the presence of Na-O and C-N anticorrelations. Fe- and $s$-poor stars have, on average, lower sodium and nitrogen abundance than Fe- and $s$-rich ones.
Indeed, in this cluster, 
the chemical tracing of multiple stellar populations along the HB must take into account not only light elements, but also $s$-process elements.
All our analysed HB stars are consistent with having all the same Ba and Na abundances, suggesting that
they are consistent with being the progenie of the Na-poor and $s$-poor first generation stars observed along the RGB.

Similarly to what recently observed in {\it normal} GCs, the reddest HB stars of M\, 22 belong to the primordial population, whose light-element chemical composition (\eg\ Na) is consistent with field stars at similar metallicity.
As a consequence of this, it is tempting to speculate that He could be an important parameter in determining the HB morphology in GCs, as second generation(s) stars are supposed to born from H-burning processed material.   
He-rich stars should evolve faster than He-poor ones and hence, present-day Na-rich stars should be less massive than Na-poor stars. 
As a consequence of their mass, He/Na-rich and He-Na-poor stars should evolve into different HB regions, with Na-rich HB stars having also redder colors.
However, direct and strong observational proofs supporting this idea are still missing.

Some indication for a possible He enhancement of second generation(s) stars have been shown by Dupree et al. (2011) and Pasquini et al. (2011), on the basis of the detection of a chromospheric He line for Na-rich RGB stars.
In addition to this, the comparison between the observed multiple sequences along the CMDs and theoretical models of a number of GCs (e.g.\ Piotto et al. 2007, Milone et al.\ 2012a) suggest that present-day Na-rich stars in GCs could be enhanced in He.
On the other hand, the recent results of Moni Bidin et al. (2011) cast many doubts, as
they show that blue HB stars in $\omega$~Centauri are not brighter than in other GCs, as would be expected as a consequence of a possible He enhancement.

In M\, 22 a further complication comes from the presence of star-to-star variations in the overall C+N+O content. Theoretical models, indeed, predict that the effective temperature of stars along the HB, as well as the mass loss along the RGB is affected by the CNO abundance (Cassisi et al.\ 2008, Ventura et al.\ 2009). 
The fact that 
all the five targets for which barium content has been measured show abundances consistent with the $s$-poor group identified on the RGB (Marino et al. 2009), showing C+N+O abundances lower than in $s$-rich stars,
supports the hypothesis that the CNO abundance can play a role in determining the distribution of stars along the HB.  
In general, the result that a small HB segment is populated by stars with almost the same chemical composition, supports the idea that the presence of multiple stellar generations with different chemical composition is a key parameter in determining the HB morphology.

%__________________________________________________________________
\begin{acknowledgements}
We warmly thank the anonymous referee for his/her comments, that helped to strengthen our results.
We are grateful to Peter Stetson for providing photometry of M\,22. 
APM acknowledges the financial support from the Australian Research 
Council through Discovery Project grant DP120100475.
Support for this work has been provided by the IAC (grant 310394), 
and the Education and Science Ministry of Spain (grants AYA2007-3E3506, and AYA2010-16717).
\end{acknowledgements}
%__________________________________________________________________
\bibliographystyle{aa}

\begin{table*}[ht!]
\scriptsize
\begin{center}  
\caption{Coordinates, radial velocities RV ([\kmsec]), projected rotational velocities \vsini\  ([\kmsec]), and atmospheric parameters for the target stars.}
\begin{tabular}{lcccccccccccccc}
\hline
\hline
ID       &$\alpha$(J2000)&$\delta$(J2000)& RV                         &\vsini&\teff&\logg&\vmicro\tablenotemark{a}&&\teff&\logg&[A/H]&\vmicro&&\logg \\
         &                         &                        &                               &      &\multicolumn{3}{c}{isochrones}&&\multicolumn{4}{c}{adopted}&&H$_{\alpha}$ \\
\hline

166      &18:36:07.73 &$-$23:55:43.4     &$-$138.8   &   15  & 7970 & 3.10 & 2.56 & &7700 & 3.95 & $-$1.76&1.50&&3.30 \\
2212    &18:36:30.36 &$-$23:57:12.7     &$-$153.0   &   31  & 7643 & 3.02 & 2.65 & &7550 & 3.02 & $-$1.55&2.67&&2.69 \\
2297    &18:36:31.12 &$-$23:51:45.1     &$-$142.3   &   17  & 8317 & 3.17 & 2.48 & &7850 & 3.17 & $-$1.83&2.50&&3.02 \\
2346    &18:36:31.66 &$-$23:49:30.2     &$-$144.5   &   17  & 7445 & 2.99 & 2.70 & &7400 & 2.99 & $-$1.63&2.40&&2.81 \\
2548    &18:36:36.96 &$-$23:54:33.1     &$-$142.4   &   16  & 7366 & 2.97 & 2.72 & &7330 & 2.97 & $-$1.70&2.10&&2.71 \\
2562    &18:36:37.68 &$-$23:57:24.3     &$-$136.1   &   26  & 8317 & 3.20 & 2.48 & &8317 & 3.20 & $-$1.61&2.10&&3.36 \\
  90      &18:36:02.44 &$-$23:53:38.2     &$-$148.2   &   12  & 8227 & 3.19 & 2.50 & &8000 & 3.19 & $-$1.67&2.10&&3.13 \\
\hline
\label{data}
\end{tabular}
\end{center}
\tablenotetext{a}{From the Pace at al. (2006) empirical relation, by assuming temperatures from isochrones.}
\end{table*}

\begin{table*}[ht!]
\scriptsize
\begin{center}  
\caption{Derived chemical abundances for the M22 HB stars, and mean RGB stars abundances from Marino et al. (2009, 2011) in the entire, the $s$-poor and the $s$-rich samples.}
\begin{tabular}{lccccccccccc}
\hline
\hline
ID       &[Fe/H]$_{\rm I}$\tablenotemark{a}&[Fe/H]$_{\rm I}$&[Fe/H]$_{\rm II}$&[Fe/H]$_{\rm II}$& [Na/Fe] & [Na/Fe] &[Mg/Fe]&[Ca/Fe]&[Ti/Fe]$_{\rm II}$  &[Cr/Fe]&[Ba/Fe]        \\
         &LTE &NLTE&LTE&NLTE&LTE&NLTE&LTE&LTE&LTE&LTE&LTE\\
\hline
\multicolumn{12}{c}{HB stars}\\
166    &$-$1.89 &$-$1.74&$-$1.75&$-$1.76&$-$0.11   &$-$0.39   &$+$0.22  &$+$0.46&$+$0.52&$+$0.14 & $+$0.14\\     
2212  &$-$1.79 &$-$1.51&$-$1.55&$-$1.51&$+$0.15  &$-$0.32   & $+$0.28 &$+$0.47&$+$0.37 &$-$0.06 & $-$0.20\\
2297  &$-$2.05 &$-$1.78&$-$1.83&$-$1.80&$-$0.07   &$-$0.43  & $+$0.34 &\nodata &$+$0.36 &\nodata & \nodata\\
2346  &$-$1.87 &$-$1.59&$-$1.61&$-$1.58&$+$0.24  &$-$0.26   &$+$0.30 &$+$0.46&$+$0.35&$-$0.13 &$-$0.40 \\
2548  &$-$1.94 &$-$1.67&$-$1.70&$-$1.68&$+$0.12  &$-$0.35   &$+$0.39 &$+$0.39&$+$0.30 &$-$0.23 &$-$0.35\\
2562  &\nodata &\nodata &$-$1.61&$-$1.58&$+$0.10  &$-$0.31   &$+$0.42 &\nodata &$+$0.39 &\nodata &  \nodata\\  
  90   &$-$1.86  &$-$1.65&$-$1.67&$-$1.65&$-$0.08  &$-$0.39&$+$0.34&\nodata &$+$0.32 &\nodata&$\leq -$0.05\\
\hline
avg.    &$-$1.90 &$-$1.64 &$-$1.66 &$-$1.63&$+$0.08  &$-$0.34  & $+$0.35 & $+$0.44 &$+$0.35  &$-$0.14 & $-$0.32  \\
$\pm$&\phantom{+}0.05&\phantom{+}0.05&\phantom{+}0.04&\phantom{+}0.05&\phantom{+}0.06&\phantom{+}0.03&\phantom{+}0.02&\phantom{+}0.03&\phantom{+}0.01&\phantom{+}0.06&\phantom{+}0.07 \\
$\sigma_{\rm   obs}$&\phantom{+}0.10&\phantom{+}0.10&\phantom{+}0.10&\phantom{+}0.10&\phantom{+}0.13&\phantom{+}0.06&\phantom{+}0.05&\phantom{+}0.04&\phantom{+}0.03&\phantom{+}0.09&\phantom{+}0.10\\
\hline
\multicolumn{12}{c}{RGB all stars}\\
avg.    &$-$1.76 &\nodata&$-$1.76 &\nodata&$+$0.26  &$+$0.17  & $+$0.39 & $+$0.30 &$+$0.32  &$-$0.13 & $+$0.09  \\
$\pm$&\phantom{+}0.02&\nodata&\phantom{+}0.02&\nodata&\phantom{+}0.05&\phantom{+}0.04&\phantom{+}0.02&\phantom{+}0.01&\phantom{+}0.01&\phantom{+}0.02&\phantom{+}0.04 \\
$\sigma_{\rm  obs}$&\phantom{+}0.10&\nodata&\phantom{+}0.10&\phantom{+}&\phantom{+}0.27&\phantom{+}0.25&\phantom{+}0.11&\phantom{+}0.07&\phantom{+}0.07&\phantom{+}0.08&\phantom{+}0.22\\
\multicolumn{12}{c}{RGB $s$-poor (Na-poor$+$Na-rich)}\\
avg.    &$-$1.82 &\nodata &$-$1.82 &\nodata&$+$0.15  &$+$0.08  & $+$0.38 & $+$0.30 &$+$0.32  &$-$0.16 & $-$0.05  \\
$\pm$&\phantom{+}0.02&\nodata&\phantom{+}0.02&\nodata&\phantom{+}0.06&\phantom{+}0.05&\phantom{+}0.03&\phantom{+}0.01&\phantom{+}0.01&\phantom{+}0.02&\phantom{+}0.03 \\
$\sigma_{\rm obs}$&\phantom{+}0.07&\nodata&\phantom{+}0.07&\nodata&\phantom{+}0.26&\phantom{+}0.24&\phantom{+}0.12&\phantom{+}0.04&\phantom{+}0.06&\phantom{+}0.05&\phantom{+}0.12\\
\multicolumn{12}{c}{RGB $s$-rich (Na-poor$+$Na-rich)}\\
avg.    &$-$1.67 &\nodata &$-$1.67 &\nodata&$+$0.42  &$+$0.31  & $+$0.39 & $+$0.36 &$+$0.32  &$-$0.09 & $+$0.31  \\
$\pm$&\phantom{+}0.01&\nodata&\phantom{+}0.01&\nodata&\phantom{+}0.06&\phantom{+}0.05&\phantom{+}0.03&\phantom{+}0.02&\phantom{+}0.02&\phantom{+}0.04&\phantom{+}0.04 \\
$\sigma_{\rm obs}$&\phantom{+}0.05&\nodata&\phantom{+}0.05&\nodata&\phantom{+}0.20&\phantom{+}0.19&\phantom{+}0.11&\phantom{+}0.06&\phantom{+}0.06&\phantom{+}0.08&\phantom{+}0.13\\
\hline
\label{abundances}
\end{tabular}
\end{center}
\tablenotetext{a}{The mean values for HB stars have been calculated excluding the star \#166.}
\end{table*}

\begin{table*}[ht!]
\caption{Sensitivity of derived abundances to the atmospheric parameters and EWs. We report the variations of chemical abundances due to the atmospheric parameters, errors in EW measurements ($\sigma _{\rm EW}$), and the squared sum of the internal errors ($\sigma _{\rm total}$).
\label{errors}}
\begin{tabular}{lcccccccccc}
\hline\hline
                                   &$\Delta$\teff &$\Delta$\logg&$\Delta$\vmicro&$\Delta$[A/H]& $\sigma _{\rm EW}$&$\sigma_{\rm total}$&&$\Delta$\teff &$\Delta$\logg &$\Delta$\vmicro\\  
                                &   $\pm$170~K&  $\pm$0.02  &   $\pm$0.50  &  $\pm$0.10  & $\pm$8 m\AA  &   & &$\pm$200~K   &  
$\pm$0.15 &$\pm$0.30 \\
                                 &\multicolumn{6}{c}{internal}&&\multicolumn{3}{c}{systematic}\\ \hline
$\rm {[Na/Fe]_{I}}$      &     $\mp$0.01  &  $\pm$0.00  & $\mp$0.04   & $\pm$0.00   & $\mp$0.08      &0.09 & &$\mp$0.01 & $\pm$0.00&$\mp$0.03 \\  
$\rm {[Mg/Fe]_{I}}$ 	  &     $\mp$0.03  &  $\pm$0.00  & $\mp$0.06   & $\pm$0.00   & $\mp$0.06      &0.09 & &$\mp$0.03 & $\pm$0.00&$\mp$0.04\\  
$\rm {[Ca/Fe]_{I}}$ 	  &     $\mp$0.13  &  $\pm$0.00  & $\pm$0.02   & $\pm$0.00   & $\mp$0.11      &0.17 & &$\mp$0.16 & $\pm$0.02&$\pm$0.01\\  
$\rm {[Ti/Fe]_{II}}$ 	  &     $\pm$0.02  &  $\mp$0.00  & $\pm$0.10   & $\pm$0.00   & $\pm$0.02      &0.10 & &$\pm$0.03 & $\pm$0.00&$\pm$0.06\\  
$\rm {[Cr/Fe]_{I}}$ 	  &     $\mp$0.13  &  $\pm$0.00  & $\pm$0.02   & $\pm$0.00   & $\mp$0.16      &0.21 & &$\mp$0.16 & $\pm$0.02&$\pm$0.01\\  
$\rm {[Fe/H]_{I}}$  	  &     $\pm$0.14  &  $\mp$0.00  & $\mp$0.01   & $\pm$0.00   & $\pm$0.06      &0.15 & &$\pm$0.17 & $\mp$0.02&$\mp$0.00\\  
$\rm {[Fe/H]_{II}}$ 	  &     $\pm$0.06  &  $\pm$0.01  & $\mp$0.12   & $\pm$0.00   & $\pm$0.08      &0.16 & &$\pm$0.08 & $\pm$0.04& $\mp$0.08\\  
$\rm {[Ba/Fe]_{II}}$	  &     $\pm$0.16  &  $\pm$0.00  & $\pm$0.00   & $\mp$0.10   & $\mp$0.13\tablenotemark{a}&0.23&&$\pm$0.19&$\pm$0.02&$\pm$0.00\\ 
\hline
\end{tabular}
\tablenotetext{a}{Uncertainty introduced by the continuum placement.}
\end{table*}

%__________________________________________________________________
%__________________________________________________________________

%__________________________________________________________________
%__________________________________________________________________

%__________________________________________________________________
%__________________________________________________________________

\end{document}